\documentclass[referee]{raa}
\usepackage{multirow}
\usepackage{graphicx,times}
\usepackage{subcaption}
\usepackage{natbib}
\usepackage{enumitem}
\usepackage{booktabs}
\usepackage{amssymb,amsmath}
\usepackage{hyperref}
\usepackage{caption}
\usepackage{xcolor}
\usepackage{ulem}

\bibpunct{(}{)}{;}{a}{}{,}
\usepackage{hyperref}

\usepackage[a4paper, margin=1in]{geometry}

\begin{document}

   \title{Testing the phenomenological interacting dark energy model with gamma-ray bursts and Pantheon+ type Ia supernovae}

 \volnopage{ {\bf 20XX} Vol.\ {\bf X} No. {\bf XX}, 000--000}
   \setcounter{page}{1}

   \author{Xiao-Dong Nong
   \inst{1,2}, Nan Liang\inst{1,2,3}
   }

   \institute{Key Laboratory of Information and Computing Science Guizhou Province,
   	Guizhou Normal University, Guiyang, Guizhou 550025, China; {\it liangn@bnu.edu.cn}\\
        \and
             School of Cyber Science and Technology, Guizhou Normal University, Guiyang, Guizhou 550025, China;\\
        \and
        Joint Center for FAST Sciences Guizhou Normal University Node, Guiyang, Guizhou 550025, China\\
\vs \no
   {\small Received 20XX Month Day; accepted 20XX Month Day}
}

\abstract{In this paper, we utilize recent observational data from gamma-ray bursts (GRBs) and  Pantheon+ supernovae Ia (SNe Ia) sample to explore the interacting dark energy (IDE) model in a phenomenological scenario.
Results from GRBs alone, SNe Ia and GRBs+SNe Ia indicate that the energy is transferred from dark energy to dark matter and the coincidence problem is alleviated. The value of $H_0$ from GRBs+SNe Ia in the IDE scenario shows agreement with the SH0ES measurement. Considering the age estimate of the quasar APM 08279+5255 at $z = 3.91$, we find that the phenomenological IDE scenario can predict a cosmic age greater than that of the $\Lambda$CDM model, thus the cosmic age problem can be alleviated.
}

   \authorrunning{Nong \& Liang }            
   \titlerunning{Testing IDE model with GRBs and SNe Ia}  
   \maketitle

%
\section{Introduction}           
\label{sect: intro}
Observations of type Ia Supernovae (SNe Ia)  have unequivocally demonstrated that our universe is experiencing accelerated expansion \citep{Riess1998, Perlmutter1999, Astier2006, Hicken2009, Amanullah2010}. Additional supports come from cosmic microwave background (CMB) radiation (\citealt{Spergel2003, Spergel2007, Komatsu2009, Komatsu2011}) and large-scale structure (LSS)  (\citealt{Tegmark2004, Eisenstein2005, Percival2010}).
However, the standard $\Lambda$CDM model which assumes dark energy (DE) is a cosmological constant $\Lambda$ 
encounters the cosmological constant problem \citep{Weinberg1989, Carroll1992}, encompassing the fine-tuning and cosmic coincidence issues. 
Moreover, the $H_0$ tension with the discrepancy between local measurements by SNe Ia \citep{Riess2006, Riess2018, Riess2019, Riess2022a, Riess2022b} and early Universe measurements from CMB observations \citep{Planck2016, Planck2020} under the $\Lambda$CDM model exceeding $5\sigma$ represents a significant challenge in modern cosmology.


To address or mitigate the coincidence problem, many alternative dynamic models of DE have been proposed \citep{Ratra1988, Caldwell1998, Caldwell2002, Caldwell2003, Armendariz2001, Bento2002, Chiba2002, Feng2005}. 
The interacting dark energy (IDE) models, which involve potential interactions between DE and dark matter (DM), have also been proposed and investigated to alleviate the coincidence problem \citep{Amendola2000, Zimdahl2001, Chimento2003, Guo2005, Guo2005b, WeiZhang2007a, WeiZhang2007b, Cai2010, Li2014, Goswami2019}.
Many studies suggest that interactions between DE and DM can lead to an increased value of $H_0$, potentially alleviating the tension between measurements from the Planck collaboration (within the $\Lambda$CDM paradigm) and the SH0ES collaboration \citep{Kumar2016, Kumar2017, Kumar2019,  Arevalo2017, Gonzalez2018, vonMarttens2019, Valentino2017, Yang2018, Pan2019, Pan2020, Yang2020}.
Recently, \cite{Cheng2021} investigated the possible interaction between DE and weakly interacting massive particle (WIMP) DM using observational data in cosmology.
\cite{Wang2022} found that the combined data from Pantheon+, CMB, baryon acoustic oscillations (BAOs), and the observational Hubble data (OHD) suggest energy transfer from DE to DM at  1.85$\sigma$ confidence level;
\cite{Zhao2023} used the mock Fast radio bursts (FRBs) to constrain the dimensionless coupling parameter 
in four phenomenological IDE models;
\cite{Hou2023}  constrained four phenomenological IDE models using the joint observation of gravitational wave (GW) and gamma-ray bursts;
\cite{LiZhang2023} developed a full numerical routine to solve the background and perturbation equations of the IDE models.
\cite{Li2024} simulated  GW data of four detection strategies 
to perform cosmological analysis in four phenomenological IDE models.
\cite{Halder2024} considered interaction scenarios that depend on the intrinsic nature of DE and DM, extending the unidirectional energy flow assumption to allow for bidirectional energy flow with sign shifting interaction functions.


The IDE models can be explored phenomenologically by considering the ratio of the energy densities of dark energy to matter 
as $r \equiv \rho_{X} \propto \rho_{m}a^{\xi}$, where the variable $\xi$ measures the severity of the coincidence problem \citep{Dalal2001}. 
The interaction term can be obtained by $Q=-H\rho_m(\xi+3w_X)$ in a flat FLRW universe, where $w_{\rm X}$ represents the equation of state (EoS) for DE. The condition $\xi+3w_X\neq 0$ denotes the interacting scenario.
 \cite{Chen2010} constrained the phenomenological model utilizing SNe Ia, CMB, and BAO;
\cite{Cao2011} investigated observational constraints for this phenomenological interacting scenario by integrating OHD with the joint data. 
More recently, \cite{Zheng2022} found that the simulated GW can decrease the uncertainty associated with $H_0$ 
from the joint sample in the frameworks of two typical dynamical models: the $w_0$$w_a$CDM model and this phenomenological model. 

Another possible difficulty for the $\Lambda$CDM model is related to the age problem. This issue arises from observations of certain astrophysical objects: the quasar APM 08279+5255 at $z = 3.91$ with  a rough estimate of $t_{\text{QSO}} = 2.0 \sim 3.0$ Gyr \citep{Hasinger2002}, and 1$\sigma$ lower age limit 1.8 Gyr \citep{Frica2005}, which appear to be older than the universe itself according to the $\Lambda$CDM model.
The cosmic age problem relative to the quasar APM 08279+5255 has been investigated in various studies \citep{Sethi2005,WeiZhang2007c, YangZhang2010, CuiZhang2010, Chimento2013, YuWang2014, Yan2015}.

The cosmic age problem of quasar APM 08279+5255 can also be investigated in interacting scenarios. \cite{WangZhang2008} showed that introducing dark energy alone does not solve the cosmic age problem when considering the quasar APM 08279+5255, suggesting that the potential interaction between dark energy and dark matter could offer a solution.
\cite{Wang2010} 
found that 5 globular clusters and the quasar exhibit an age discrepancy with the $\Lambda$CDM model at a confidence level exceeding 2$\sigma$, even when considering observational constraints from SNe Ia, BAO, CMB, and $H_0$ measurements; considering IDE models can extend the cosmic age, however, IDE models still struggle to fully resolve the age problem.
More recently,
\cite{ZarandiEbrahimi2022} discussed cosmic age problem within ghost dark energy (GDE) model in the presence of three types of the sign-changeable interaction terms.

Observational datasets have played a crucial role in constraining cosmology. SNe Ia data spans only up to a redshift of approximately $z \sim 2$, while CMB data pertains to redshifts near $z \sim 1100$. Therefore, cosmological data in the intermediate region 
could potentially provide vital understandings of the origins of the coincidence problem, $H_0$ tension, and the cosmic age problem.
Gamma-ray bursts (GRBs) are known to reach a higher maximum observable redshift, approximately $z \sim 10$ \citep{Cucchiara2011}. 
Due to the scarcity of samples at low redshift, a fiducial cosmological model was necessary to presuppose for the calibration of the GRBs luminosity relation in the early research \citep{Dai2004, Schaefer2007}.
\cite{Liang2008} introduced a cosmological model-independent method for calibrating the luminosity relations of GRBs by utilizing data from SNe Ia.
On the other hand, the simultaneous method \citep{Amati2008} has been also put forward to avoid the circularity problem by fitting the parameters of the correlation and cosmological models simultaneously.
\cite{Amati2019} proposed an alternative approach for calibrating the GRB relation  
using OHD by the cosmic chronometers (CC) method. 
Therefore, GRBs data can be employed to place constraints on cosmological models without the circularity problem \footnote{For a comprehensive review of the circularity problem in GRB cosmology, see \cite{Wang2015}.} \citep{Wei2010a, Wei2010b, Liang2010, Liang2011, Wang2016, Demianski2017a, Demianski2017b, Dainotti2017, Dainotti2018, Dirirsa2019, Luongo2020, Luongo2021a, Luongo2021b, Montiel2021, Liu2022, Dainotti2023}.
It should be noted that the classification of GRBs is important for the calibration of GRB relations. If GRBs can be treated as standard candles, only a small fraction of GRBs can be suitable for cosmological use. See e.g. \cite{WH2022} and \cite{Hu2021} for recent progress focused on specific GRB types to improve their standardization and enhance their utility in cosmology.

More recently,
\cite{Khadka2021} made a compilation of a dataset comprising 118 GRBs with the smallest intrinsic dispersion from a broader set of 220 GRBs. 
\cite{Liang2022} and \cite{Li2023} calibrated these GRB data using a Gaussian Process,
from Pantheon sample \citep{Scolnic2018} or the most recent OHD by the CC method \citep{Moresco2022}.
\cite{WLL2024} investigate the phenomenologically emergent dark energy (PEDE) model with GRBs and OHD at intermediate redshift.
In this paper, our goal is to test the phenomenological interacting scenario \footnote{In this work, we use $\xi$IDE to represent this phenomenological interaction model.}
from cosmology-independent gamma-ray bursts (GRBs) at redshifts ranging from 1.4 to 8.2 \citep{Li2023} and the Pantheon+ sample,
which is the most recent compilation of SNe Ia including 1701 light curves from 1550 unique data at redshifts ranging from 0.001 to 2.26 \citep{Scolnic2022}. To further explore the cosmic age problem, we incorporate the quasar APM 08279+5255 at redshift $z=3.91$ as an extra data point.
The rest of the paper is organized as follows: Sec. \ref{sect: model} provides an overview of the phenomenological IDE model. The observational data adopted in this study are described in Sec. \ref{sect: data}. Constraints from the observational data are given in Sec. \ref{sect: results}. We give a brief conclusion in the Sec. \ref{sect: con}.

\section{Overview of the phenomenological interacting model}
\label{sect: model}
In this paper, we explore the possibility of DE and matter engage in an energy exchange through an interaction term denoted as $Q$ 
\begin{equation}
	\dot{\rho}_{X} + 3H\rho_{X}(1 + w_{\rm X}) = -Q, \nonumber\\
\end{equation}
\begin{equation}
	\dot{\rho}_m + 3H\rho_m = Q.
\end{equation}
The phenomenological model adopts an assumption that the ratio of the dark energy to matter densities is \citep{Dalal2001}
\begin{equation}
	\rho_{\rm X} \propto \rho_{\rm m}a^{\xi},
\end{equation}
values of $\xi = 3$ and $\xi = 0$ are associated with the $\Lambda$CDM model and a self-similar solution that is free from the coincidence problem, respectively; while $0 < \xi \leq 3$ indicates a less severe coincidence problem \citep{Pavon2004}.
Considering the phenomenological 
model in a flat FLRW universe with $\Omega_{\rm X0} + \Omega_{\rm m0} = 1$, we can derive  the interaction term,
\begin{equation}
	Q = -H\rho_{\rm m}(\xi +3w_{\rm X})\Omega_{X},
\end{equation}
where $\Omega_{X} = \frac{1-\Omega_{\rm m0}}{(1-\Omega_{\rm m0})+ \Omega_{\rm m0}(1+z)^\xi}$, the case $\xi+3w_{\rm X}=0$ (which implies $Q = 0$) corresponds to the standard cosmology without any interaction between dark energy and matter.
Conversely, $\xi+3w_{\rm X} \neq 0$ indicates the non-standard cosmology.
Furthermore, when $\xi+3w_{\rm X} > 0$ ($Q < 0$), it suggests that energy is transferred from matter to DE, which tends to exacerbate the coincidence problem.
Oppositely, if $\xi+3w_{\rm X} < 0$ ($Q > 0$), it indicates that the energy is transferred from DE to DM, which could potentially  mitigate the coincidence problem.
The $H(z)$ function of the phenomenological model can be expressed as \citep{Cao2011}
\begin{eqnarray}\label{eq8}
	H(z)
	= H_0 \sqrt{(1+z)^{3}\left[\Omega_{m0}+
		(1-\Omega_{m0})(1+z)^{-\xi}\right]^{-3w/\xi}}.
\end{eqnarray}

For comparison, we also utilize the Chevallier-Polarski-Linder (CPL) model \citep{Chevallier2001, Linder2003}, where dark energy is represented as evolving with redshift through the parametrization $w=w_0+w_az/(1+z)$.
The $\Lambda$CDM model corresponds to $w_0=-1, w_a=0$. 
The $H(z)$ function of them can be expressed as
\begin{eqnarray}\label{eq9}
	H(z)
	= H_0 \sqrt{\Omega_{m0}(1+z)^{3} + (1-\Omega_{m0})(1+z)^{3(1+w_0+w_a)}e^{-\frac{3w_az}{1+z}}}.
\end{eqnarray}

To evaluate the efficacy of various models when applied to diverse data set combinations, we proceed to quantify the variances in the Akaike information criterion (AIC; \cite{Akaike1974, Akaike1981}) and the Bayesian information criterion (BIC; \cite{Schwarz1978}), as defined by the following equations
\begin{gather}
	\label{eqnarray_AIC} \mathrm{AIC} = 2p - 2\ln(\mathcal{L}_{\rm max}), \\
	\label{eqnarray_BIC} \mathrm{BIC} = p\ln N - 2\ln(\mathcal{L}_{\rm max}).
\end{gather}
where $p$ is the number of parameters in the model, $N$ is the sample size of the observational data combination, and $\mathcal{L}_{\rm max}$ is the maximum likelihood of the model. The values of $\Delta$AIC and $\Delta$BIC for each model are given by AIC = $\chi^2_{min} + 2\Delta n$, and BIC = $\chi^2_{\rm min} + 2\Delta n$.
They represent the differences in $\rm AIC$ and $\rm BIC$, respectively, when compared to the values obtained for the $\Lambda$CDM model.

\section{Observational data}\label{sect: data}
The recent observations of GRB sample \citep{Khadka2021} and the latest SNe Ia sample \citep{Scolnic2022} are utilized in our cosmological analysis.
For the GRBs sample, we follow the cosmology-independent approach in \cite{Li2023} to calibrate the Amati relation using the A118 GRB sample \footnote{We use the A118 GRB sample from the A220 GRB sample used in \cite{Khadka2021} with the higher qualities appropriate to investigate cosmology.}  using the 32 updated OHD measurements. 
We utilize GRB data at $z > 1.4$ to constrain on cosmological models, the $\chi^2_{\text{GRB}}$ function is defined as:
\begin{equation}\label{eq: chi_GRB}
	\chi^2_{\text{GRB}} = \sum^{N}_{i=1} \left[\frac{\mu_{\rm obs} (z_i)-\mu_{\rm th}(z_i;p,H_0)}{\sigma_{\mu_i}}\right]^2,
\end{equation}
where $N=98$, $\mu_{\rm th}(z_i;p,H_0)$ represents the theoretical distance modulus of the model at redshifts $z_i$ , $H_0$ is the Hubble constant, $p$ denotes the cosmological parameter space,  $\mu_{\rm obs}(z_i)$ and $\sigma_{\mu_i}$ correspond to the observed value and the error, respectively.

The SNe Ia sample used in this study is the Pantheon+ sample, which comprises 1701 SNe Ia light curves observed from 1550 distinct SNe, spanning a redshift range of 0.001 to 2.26 \citep{Scolnic2022, Brout2022}.
This dataset includes major contributions from various sources, including CfA1 \citep{Riess1999}, CSP DR3 \citep{Krisciunas2017}, DES \citep{Brout2019}, PS1 \citep{Scolnic2018}, SDSS \citep{Sako2018} and SNLS \citep{Betoule2014}.
The optimal values for the cosmological parameters are derived by minimizing the $\chi^2_{\text{SNe}}$
\begin{equation}\label{eq: chi_SNe}
	\chi^2_{\text{SNe}} = \Delta\mu C_{\text{stat+syst}}^{-1} \Delta\mu^{T},
\end{equation}
where $\Delta\mu$ denotes the discrepancy between the observed distance modulus $\mu_{\rm obs}$ and its theoretical distance modulus $\mu_{\rm th}$:
\begin{equation}\label{eq: delta_mu}
	\Delta\mu = \mu_{\rm obs}(z_{i}) - \mu_{\rm th}(z_i;p,H_0).
\end{equation}
The corresponding expression for $\mu_{\rm th}$ can be given by:
\begin{equation}\label{eq: mu}
	\mu_{\rm th}(z_i;p,H_0) = m - M = 5\log_{10}\frac{d_{L}(z_i;p,H_0)}{Mpc} + 25,
\end{equation}
here, $z_{i}$ indicates the peculiar-velocity-corrected CMB-frame redshift of each SNe Ia \citep{Carr2022}, $p$ represents the cosmological parameters, $m$ denotes the apparent magnitude of the source, $M$ is the absolute magnitude, and $d_{L}$ is the luminosity distance, defined as
\begin{equation}\label{eq: dL}
	d_L = \frac{c(1+z)}{H_0} \int_{0}^{z} \frac{dz'}{H(z')},
\end{equation}
where $c$ represents the speed of light. The complete covariance matrix for Pantheon+ sample is given by \citep{Brout2022}:
\begin{equation}\label{eq: covariance}
	C_{\text{stat+syst}} = C_{\text{stat}} + C_{\text{syst}},
\end{equation}
where $C_{\text{stat}}$ and $C_{\text{syst}}$ refer to the statistical and systematic covariance matrices, respectively. The datasets and $C_{\text{stat+syst}}$ can be obtained online \footnote{\url{https://github.com/PantheonPlusSH0ES/DataRelease}}.
The total $\chi^2_{\text{total}}$ for the combined GRBs and SNe Ia data is given by $\chi^2_{\text{total}} = \chi^2_{\text{GRB}} + \chi^2_{\text{SNe}}$.
The MCMC method, facilitated by the Python package emcee \citep{ForemanMackey2013}, is employed for the minimization of the chi-square function $\chi^2$.

\section{CONSTRAINTS ON COSMOLOGICAL MODELS}\label{sect: results}

\begin{figure}[h!]
	\centering
	\includegraphics[width=0.18\textwidth]{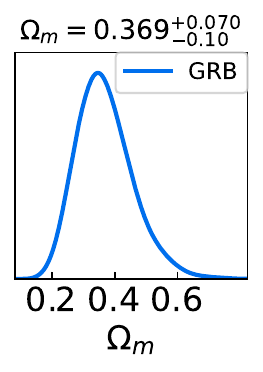}
	\includegraphics[width=0.4\textwidth]{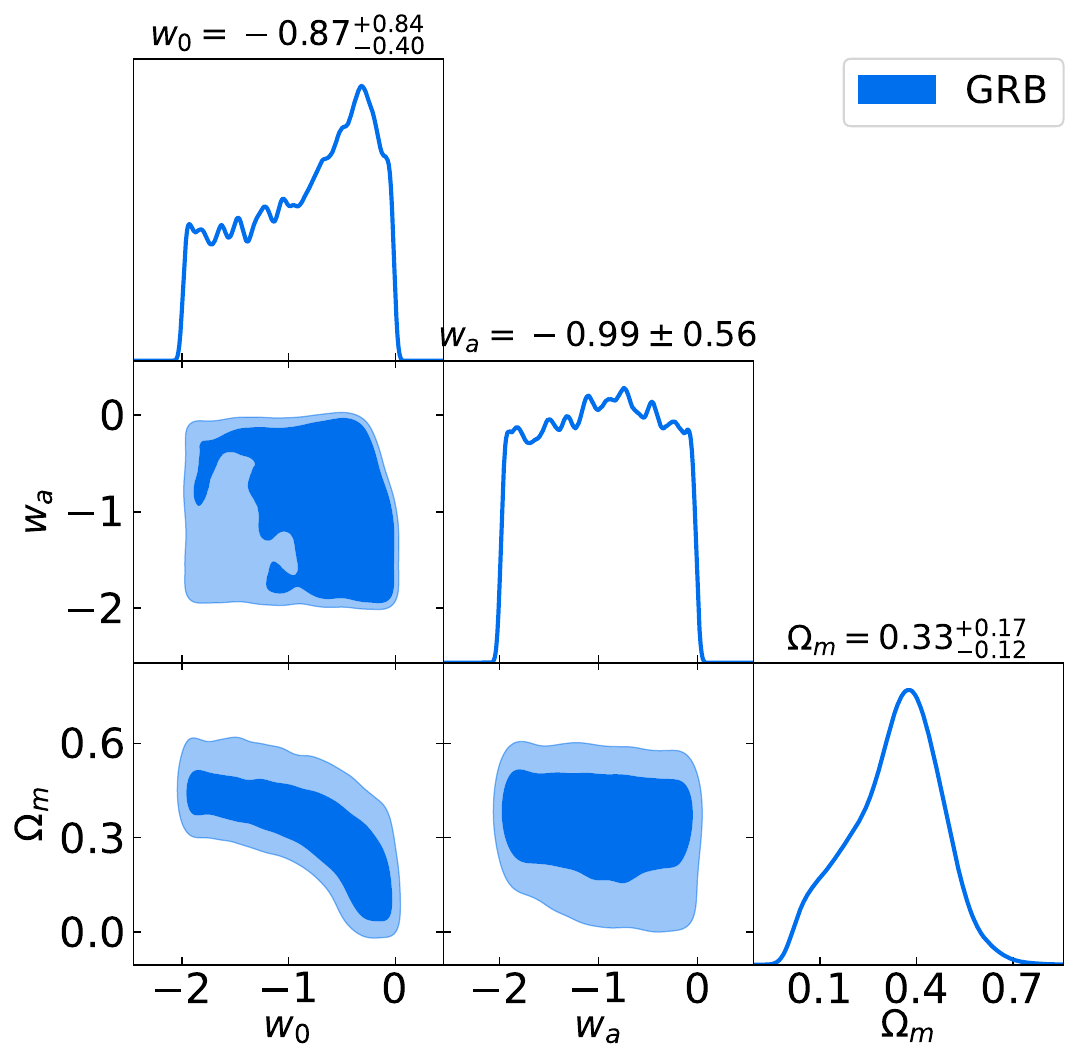}
	\includegraphics[width=0.4\textwidth]{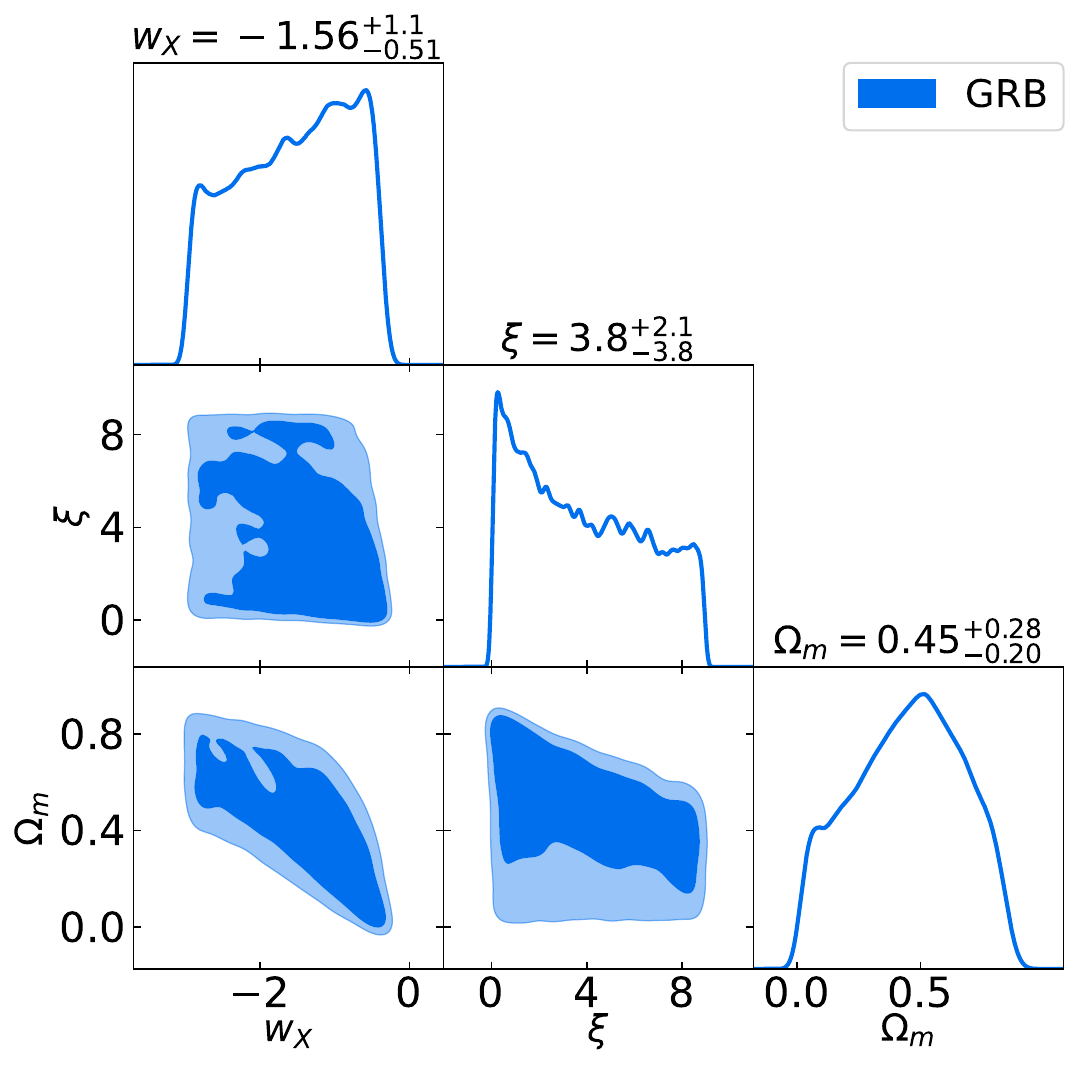}
	\caption{Constraints on the cosmological parameters of the $\Lambda$CDM model(\emph{left}), the CPL model (\emph{middle}), and the $\xi$IDE model (\emph{right}) with GRBs alone.
}
	\label{fig: models_fixed_h}
\end{figure}

Since GRB data alone cannot constrain $H_0$ due to the degeneracy with the correlation intercept parameter, we follow previous works \citep{Khadka2021, Liang2022} and fix $H_0$=$70\ {\rm km}\ {\rm s}^{-1}{\rm Mpc}^{-1}$ for the case of GRBs alone.
The results are  presented  in Fig. \ref{fig: models_fixed_h} and summarized in Tab. \ref{tab:tab1}.
For the CPL model, the constraint on the parameter $w_a$ using GRBs alone indicates a possible DE evolution ($w_a \neq 0$) within the 1$\sigma$ confidence region.
For the $\xi$IDE model,
the constraint results for $w_X$ and $\xi$
suggest: $\xi + 3w_X < 0$ with GRBs alone,
This finding is consistent with recent work \citep{Zheng2022};
however, this result is different from previous studies, which found $\xi + 3w_X > 0$ is slightly preferred 
\citep{Guo2007, Chen2010, Cao2011}.
We find our results for the $\Lambda$CDM model, $w$CDM model, and CPL model using GRBs alone are compatible with the previous works of \cite{Li2023}.

\begin{figure}[h!]
	\centering
	\includegraphics[width=0.22\textwidth]{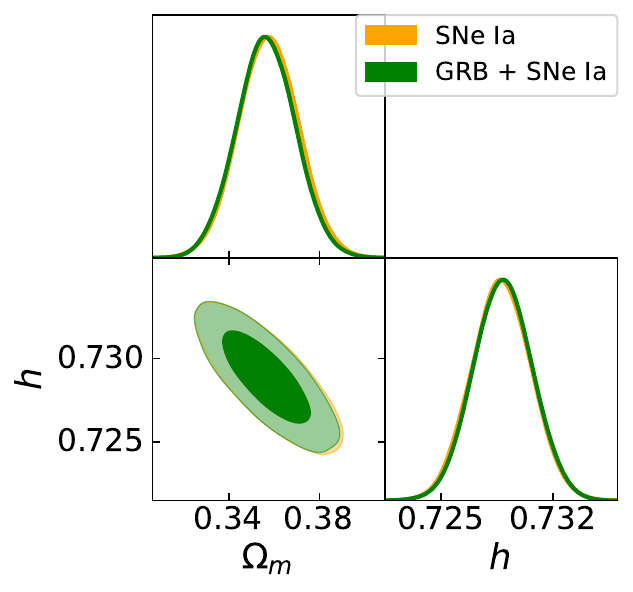}
	\includegraphics[width=0.38\textwidth]{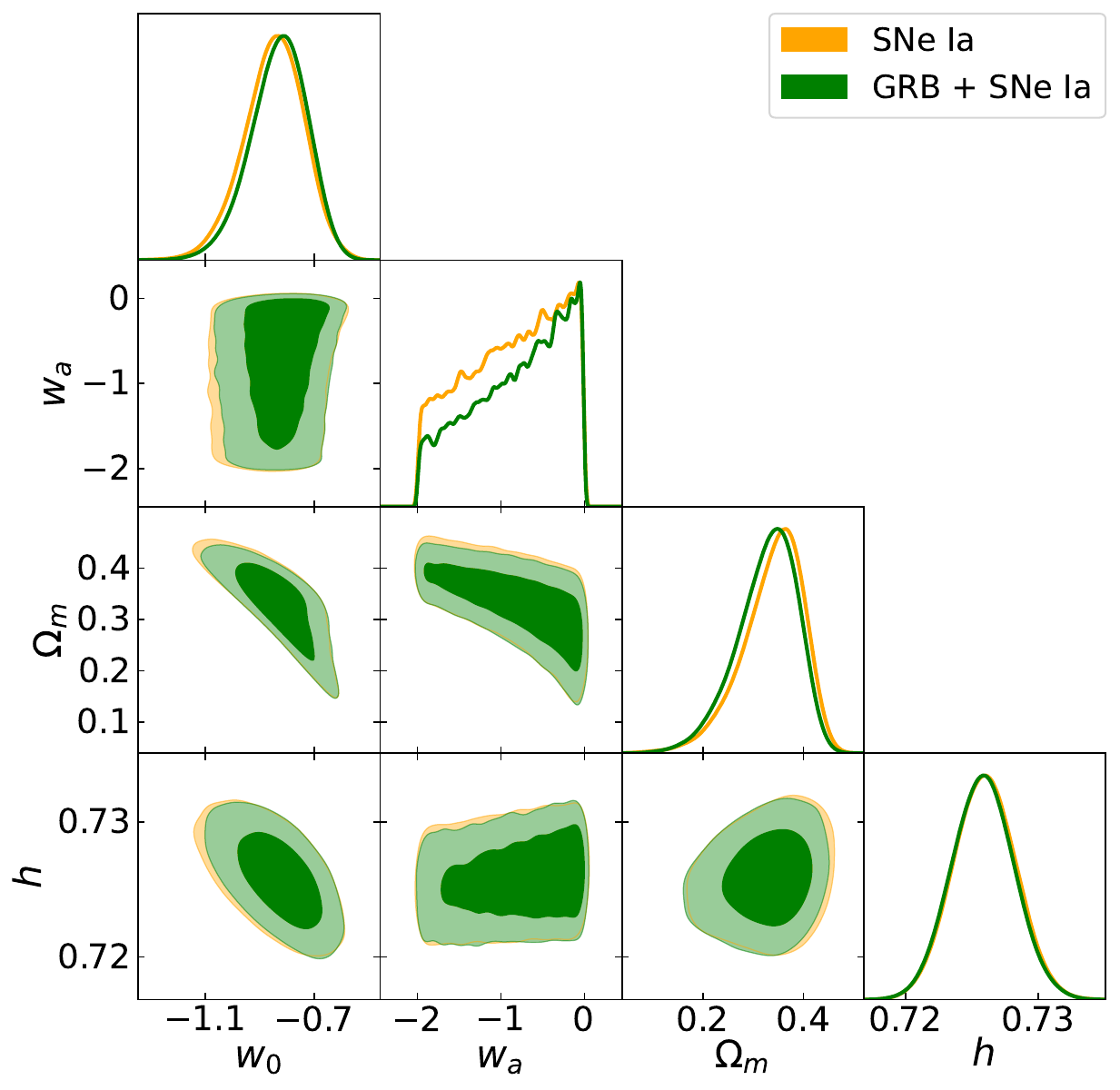}
	\includegraphics[width=0.38\textwidth]{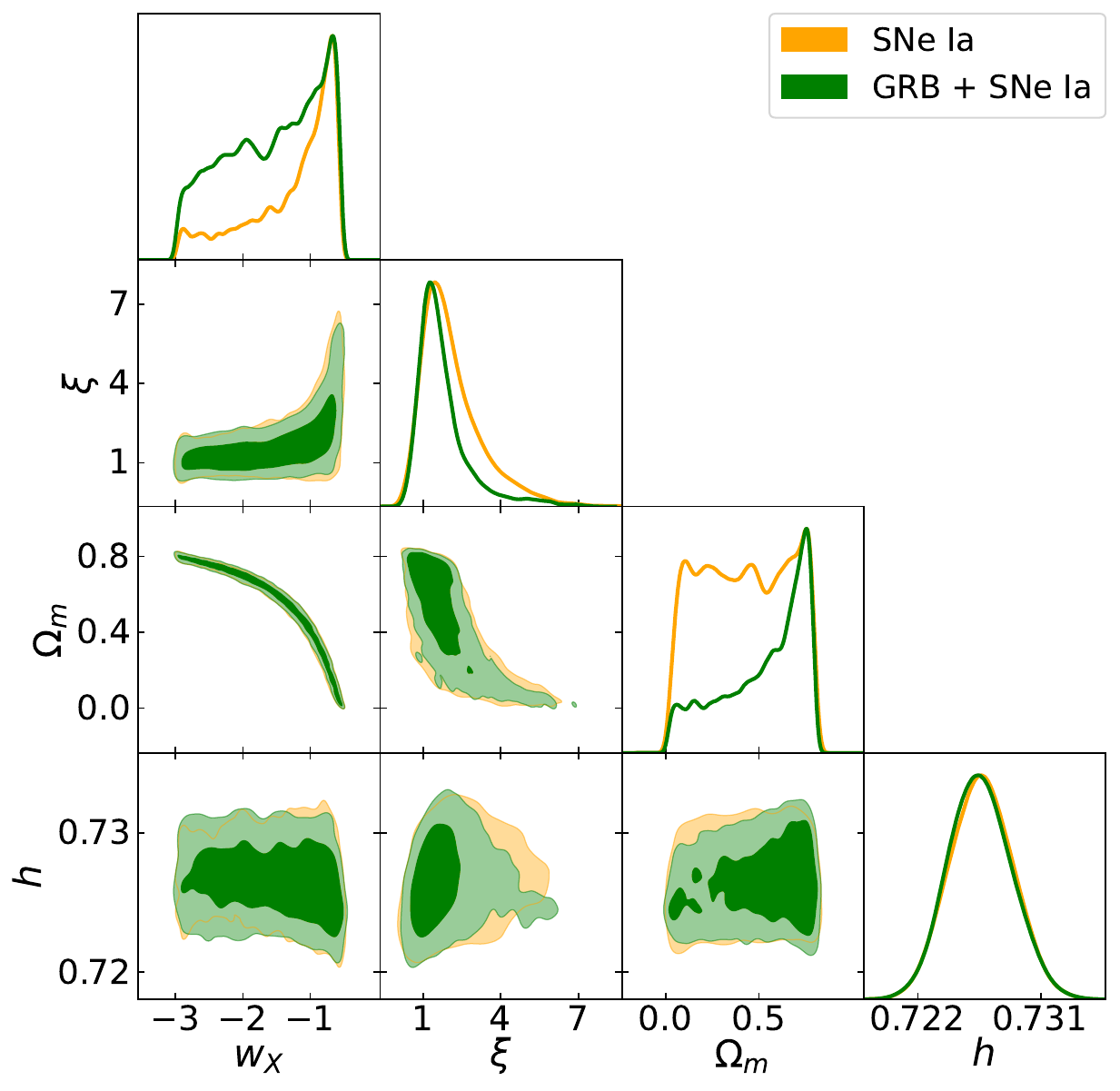}
	\caption{
	Constraints on the cosmological parameters of the $\Lambda$CDM model(\emph{left}), the CPL model (\emph{middle}), and the $\xi$IDE model (\emph{right}) with SNe Ia, and GRBs+SNe Ia.
}
	\label{fig: models}
\end{figure}

We also consider the joint constraints using GRBs and the Pantheon+ sample.
The best-fit values with the 1$\sigma$ confidence level on parameters obtained from SNe Ia, and GRBs + SNe Ia
are shown in Fig. \ref{fig: models} and summarized in Tab.\ref{tab:tab1}.
We can find that GRBs+SNe Ia is more powerful on constraining the parameters ($\Omega_m$, $h$, $w_{\rm X}$ and $\xi$) in the framework of $\Lambda$CDM, CPL and $\xi$IDE models, while GRBs-only gives weak constraints on all parameters.
We find that 
$\xi + 3w_X < 0$ with 
SNe Ia 
and GRBs+SNe Ia for the $\xi$IDE model, which is consistent with GRBs-only case. 
For the value of Hubble constant, our result with GRBs+SNe Ia for the $\xi$IDE model aligns with the one reported by SH0ES \citep{Riess2022b} ($H_0 = 73.01 \pm 0.99 {\rm\ km\ s^{-1}\ Mpc^{-1}}$) at 0.35$\sigma$.
Additionally, our result is in agreement with those results obtained in \cite{Hu2024} ($H_0=72.83\pm0.23{\rm\ km\ s^{-1}\ Mpc^{-1}}$) and \cite{Wang2022} ($H_0=73.5^{+8.1}_{-10.0}{\rm\ km\ s^{-1}\ Mpc^{-1}}$) with the 1$\sigma$ error of $h$ is reduced.
For the $\Lambda$CDM model with SNe Ia, 
our result is consistent with \cite{Brout2022} ($H_0=73.6 \pm 1.1 {\rm\ km\ s^{-1}\ Mpc^{-1}}$) and \cite{Dahiya2023} ($H_0=73.500^{+1.013}_{-0.978} {\rm\ km\ s^{-1}\ Mpc^{-1}}$), except a reduced 1$\sigma$ error for $H_0$.
This is mainly due to the application of the standardized distance modulus ($\mu_{\rm obs}=m - M$; \cite{Tripp1998}),
with $M$ being calculated from the Cepheid host galaxy distances determined by SH0ES \citep{Riess2022a}.



For the CPL model, constraints on the parameter $w_a$ derived from	
SNe Ia 
and GRBs+SNe Ia 
suggest a possible DE evolution ($w_a \neq 0$) within the 1$\sigma$ confidence region, which is  consistent with the results obtained using GRBs alone.
However, the statistical measures in Table \ref{tab:tab1} reveal that the $\xi$IDE model exhibits much higher $\rm \Delta AIC$ and $\rm \Delta BIC$ values when compared to the $\Lambda$CDM and CPL models, which indicates that the $\Lambda$CDM model is still favored.




\begin{table}
	\begin{center}
		\captionsetup{font=small}
		\caption[]{Constraints on cosmological parameters $\Omega_{\rm m}$, $h$, $w_{\rm X}$, $\xi$, $w_0$ and $w_a$ for the $\Lambda$CDM model, the CPL model, and the $\xi$IDE model using GRBs ($z > 1.4$), SNe Ia, and GRBs+SNe Ia.}
		\small
		\resizebox{\textwidth}{!}{%
			\begin{tabular}{cccccccccc}
				\hline \hline
				Parameters & $\Omega_{\rm m}$ & $h$ & $w_{\rm X}$ & $\xi$ & $w_0$ & $w_a$ & $-2\ln \mathcal{L_{\rm max}}$  & $\rm \Delta AIC$  & $\rm \Delta BIC$  \\ \hline
				GRBs &  &  &  &  &  &  &  &   &   \\ \hline
				
				

%
				
				$\Lambda$CDM &  $0.369_{-0.100}^{+0.070}$ & $0.7$ & - & - & - & - & 43.145 & - & -\\
				
				
				CPL &  $0.33_{-0.12}^{+0.17}$ & $0.7$  & - & - &  $-0.87_{-0.40}^{+0.84}$ & $-0.99\pm0.56$ & 43.212 & 4.067 & 9.238 \\

				$\xi$IDE &  $0.45^{+0.28}_{-0.20}$ & $0.7$ & $-1.56_{-0.51}^{+1.10}$ & $3.8^{+2.1}_{-3.8}$ & - & - & 43.383 & 4.238 & 9.409 \\
				\hline

				SNe Ia &  &  &  &  &  &  &  &  &     \\ \hline
				


				$\Lambda$CDM &  0.358$\pm$0.013 & 0.7288$\pm$0.0019 & - & - & - & - & 1821.892 & - & -  \\
				
				
				CPL &  $0.338_{-0.043}^{+0.072}$ & $0.7259\pm0.0024$ & - & - & $-0.849_{-0.098}^{+0.120}$ & $-0.87_{-0.32}^{+0.85}$ & 1820.629 & 2.737 & 13.615 \\
				
				$\xi$IDE & $0.42\pm0.23$ & $0.7266\pm0.0023$ & $-1.27_{-0.34}^{+0.73}$ & $2.12_{-1.40}^{+0.43}$ & - & - & 1995.103 & 177.211 & 188.089 \\
				\hline
				
				GRBs + SNe Ia &  &  &  &  &  &  &  &  &     \\ \hline
				
				

				$\Lambda$CDM &  0.357$\pm$0.013 & $0.7289\pm0.0019$  & - & - & - & - & 1865.338 & - & -\\
				
				
				CPL &  $0.326_{-0.046}^{+0.071}$ & $0.7258\pm0.0024$  & - & - & $-0.831_{-0.091}^{+0.110}$ & $-0.80_{-0.30}^{+0.79}$ & 1863.671 & 2.333 & 12.343 \\

				$\xi$IDE &   $0.525_{-0.099}^{+0.280}$ & $0.7266\pm0.0022$ & $-1.55_{-0.57}^{+1.00}$ & $1.78_{-1.10}^{+0.31}$  & - & - & 2035.557 & 243.557 & 254.501 \\
				
				\hline
		\end{tabular}}
		\label{tab:tab1}
	\end{center}
\end{table}

The age estimation of old objects at high redshifts is crucial for constraining cosmological parameters \citep{Alcaniz1999,Lima2000}.
The quasar APM 08279+5255 at redshift  $z = 3.91$ is
particularly important in this regard.
Its age has been estimated through chemical evolution studies.
Utilizing the Fe/O ratio derived from X-ray observations, \cite{Hasinger2002} provided a rough estimate of $t_{\text{QSO}} = 2.0 \sim 3.0$ Gyr. Subsequently, \cite{Frica2005} used a detailed chemodynamical model to obtain $t_{\text{QSO}} = 2.1 \pm 0.3$ Gyr at this redshift.
The age of a flat universe at redshift $z$ can be calculated by \citep{Alcaniz2003,Jain2006}:
\begin{eqnarray}\label{eq: t(z)}
	t(z) = \int_{z}^{\infty} \frac{dz'}{(1 + z') H(z')}.
\end{eqnarray}
However, the $\Lambda$CDM model predicts a cosmic age of 1.63 Gyr at $z = 3.91$ according to the 7-year WMAP data \citep{Komatsu2011} ($\Omega_m = 0.272$, $h = 0.704$), while the quasar's 1$\sigma$ lower age limit is 1.8 Gyr.

\begin{figure}[h!]
	\centering
	\includegraphics[width=0.45\textwidth]{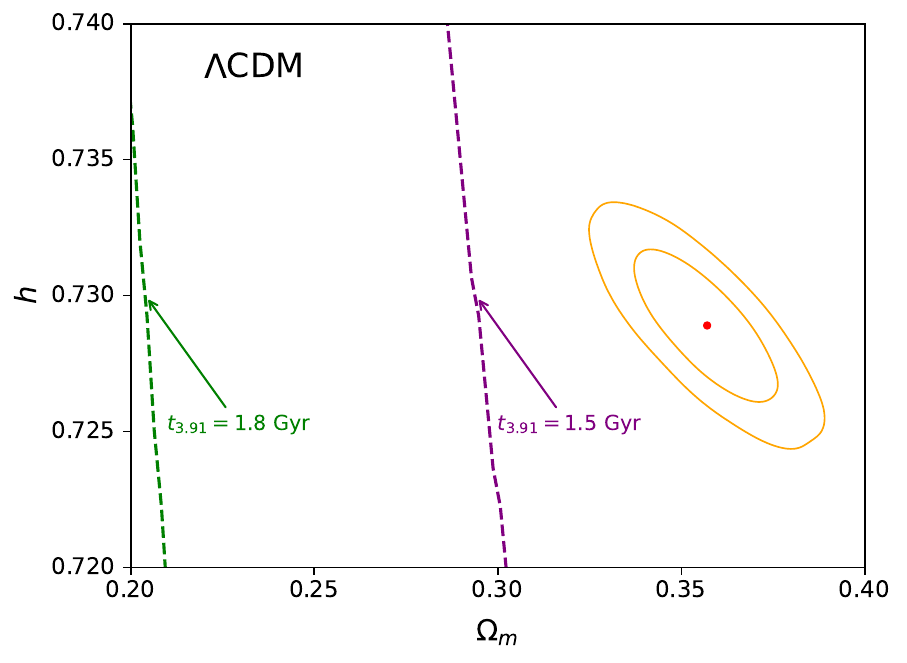}
	\hspace{0.5cm}
	\includegraphics[width=0.45\textwidth]{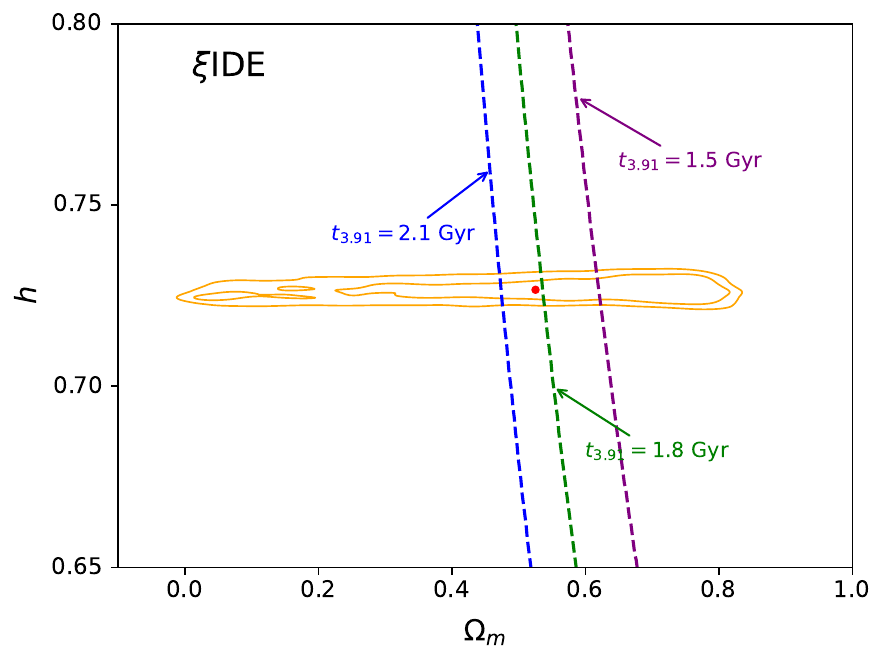}
	\caption{The contours correspond to 1$\sigma$, and 2$\sigma$ confidence regions in the $\Omega_m-h$ plane with GRBs+SNe Ia for the $\Lambda$CDM model (left panel), $\xi$IDE model (right panel), respectively. The blue, green
     and purple correspond to the cosmological parameters that would result in a universe age of 2.1 Gyr (the best-fit), 1.8 Gyr (the 1$\sigma$ limit), and 1.5 Gyr (the 2$\sigma$ limit)
    at $z = 3.91$, respectively. The central red dot denotes the bets-fit point with GRBs+SNe Ia for the $\Lambda$CDM  and $\xi$IDE models, respectively.}
	\label{fig: LCDM_Q=Q(z)_z3.91=1.8}
\end{figure}

In this work, we use the estimated lower limit ages of quasar $z = 3.91$ with the 1$\sigma$  (1.8 Gyr) and 2$\sigma$ (1.5 Gyr)
to examine the age problem in both the $\Lambda$CDM model and the $\xi$IDE model within the parameter space ($\Omega_m-h$ plane) 
using GRBs+SNe Ia data, which are shown  Fig. \ref{fig: LCDM_Q=Q(z)_z3.91=1.8}.
Our calculations indicate that the presence of
the quasar APM 08279 + 5255 is incompatible with the $\Lambda$CDM model,
which aligns with the earlier results from \cite{WangZhang2008, YangZhang2010, Wang2010}.
It can be found that the cosmic age predicted by the $\Lambda$CDM model with the best-fit values at $z = 3.91$ (1.38 Gyr) deviates 2.4$\sigma$ from the quasar APM 08279 + 5255's age (2.1 Gyr); however, the one predicted  by the $\xi$IDE model with the best-fit values at $z = 3.91$ (1.83 Gyr), which can be accommodated the quasar APM 08279 + 5255 at 0.9$\sigma$ confidence level.
Our calculations show that the $\xi$IDE model can predict significantly higher cosmic ages compared to the $\Lambda$CDM model, thereby substantially alleviating the cosmic age problem.

\section{CONCLUSION}
\label{sect: con}
In this study, we test the phenomenological interacting scenario  (the $\xi$IDE model) with the A118 GRB sample calibrated from the updated OHD measurements in a cosmology-independent approach, Pantheon+ SNe Ia sample, 
and the age lower limit of quasar APM 08279+5255 at high-$z$. 
In the scenario of the $\xi$IDE model, constraints of $\xi$ and $w_X$ derived from GRBs-only, SNe Ia-only, and GRBs+SNe Ia support $\xi + 3w_X < 0$, indicating that energy is transferred from dark energy to dark matter, which is consistent with previous results from \cite{Zheng2022}; thus the coincidence problem is alleviated.
The best-fit values of $H_0$ obtained in the $\xi$IDE model deviate by 0.35$\sigma$ from SH0ES measurements \citep{Riess2022b} using SNe Ia-only and GRBs+SNe Ia.
Considering the age estimate for the quasar APM 08279+5255, our results show that the $\xi$IDE model can accommodate this quasar, predicting significantly higher cosmic ages in comparison to the $\Lambda$CDM model and thereby substantially alleviating the cosmic age problem.



The $\Lambda$CDM model 
aligns well with most astronomical observations; however, some recent observations display slight deviations from $\Lambda$CDM model \citep{Perivolaropoulos2022, HuWang2023, Carloni2024, Colgain2024a, Colgain2024b, Colgain2024c, Luongo2024}.
Our analysis also indicates that the constraints for the CPL model with various datasets support the possibility of dark energy evolution ($w_a \neq 0$) within the 1$\sigma$ confidence region;
whereas the $\Lambda$CDM model remains favored when comparing $\rm \Delta AIC$ and $\rm \Delta BIC$ values.

Future work should focus on refining IDE models and exploring their implications with more precise observational data, such as the recent GRBs from \emph{Fermi} data \citep{WL2024} and BAOs from Dark Energy Spectroscopy Instrument (DESI) data release 1 \citep{DESI2024}. 

\begin{acknowledgements}
We very much appreciate helpful comments from anonymous referee. We are grateful to Prof. Yu Pan, Prof. Jianchao Feng, Huifeng Wang, Guangzhen Wang, Zhen Huang, and Xin Luo for useful discussions. This project was supported by the Guizhou Provincial Science and Technology Foundations (QKHJC-ZK[2021] Key 020 and QKHJC-ZK[2024] general 443). 
\end{acknowledgements}

%
\bibliographystyle{raa}
\bibliography{bibtex}

\begin{thebibliography}{136}
\providecommand\natexlab[1]{#1}
\providecommand\JournalTitle[1]{#1}

\bibitem[Akaike(1974)]{Akaike1974}
Akaike, H. 1974, ITAC, 19, 716

\bibitem[Akaike(1981)]{Akaike1981}
Akaike, H. 1981, J. Econ., 16, 3

\bibitem[Alcaniz {et~al.}(2003)]{Alcaniz2003}
Alcaniz, J.~S., Jain, D., \& Dev, A. 2003, Phys. Rev. D, 67, 043514

\bibitem[Alcaniz \& Lima(1999)]{Alcaniz1999}
Alcaniz, J.~S., \& Lima, J. A.~S. 1999, ApJ, 521, L87

\bibitem[Amanullah {et~al.}(2010)]{Amanullah2010}
Amanullah, R., Lidman, C., Rubin, D., \& et~al. 2010, ApJ, 716, 712

\bibitem[Amati {et~al.}(2008)]{Amati2008}
Amati, Guidorzi, Frontera, \& et~al. 2008, MNRAS, 391, 577

\bibitem[Amati {et~al.}(2019)]{Amati2019}
Amati, L., D'Agostino, R., Luongo, O., \& et~al. 2019, MNRAS, 486, L46

\bibitem[Amendola(2000)]{Amendola2000}
Amendola, L. 2000, Phys. Rev. D, 62, 043511

\bibitem[Arevalo {et~al.}(2017)]{Arevalo2017}
Arevalo, F., Cid, A., \& Moya, J. 2017, Eur. Phys. J. C, 77, 565

\bibitem[Armendariz-Picon {et~al.}(2001)]{Armendariz2001}
Armendariz-Picon, C., Mukhanov, V., \& Steinhardt, P.~J. 2001, Phys. Rev. D,
  63, 103510

\bibitem[Astier {et~al.}(2006)]{Astier2006}
Astier, P., Guy, J., Regnault, N., \& et~al. 2006, A\&A, 447, 31

\bibitem[Bento {et~al.}(2002)]{Bento2002}
Bento, M.~C., Bertolami, O., \& Sen, A.~A. 2002, Phys. Rev. D, 66, 043507

\bibitem[Betoule {et~al.}(2014)]{Betoule2014}
Betoule, M., Kessler, R., Guy, J., \& et~al. 2014, A\&A, 568, A22

\bibitem[Brout {et~al.}(2019)]{Brout2019}
Brout, D., Sako, M., Scolnic, D., \& et~al. 2019, ApJ, 874, 106

\bibitem[Brout {et~al.}(2022)]{Brout2022}
Brout, D., Scolnic, D., Popovic, B., \& et~al. 2022, ApJ, 938, 110

\bibitem[Cai \& Su(2010)]{Cai2010}
Cai, R.-G., \& Su, Q. 2010, Phys. Rev. D, 81, 103514

\bibitem[Caldwell(2002)]{Caldwell2002}
Caldwell, R. 2002, Phys. Lett. B, 545, 23

\bibitem[Caldwell {et~al.}(1998)]{Caldwell1998}
Caldwell, R.~R., Dave, R., \& Steinhardt, P.~J. 1998, Phys. Rev. Lett., 80,
  1582

\bibitem[Caldwell {et~al.}(2003)]{Caldwell2003}
Caldwell, R.~R., Kamionkowski, M., \& Weinberg, N.~N. 2003, Phys. Rev. Lett.,
  91

\bibitem[Cao {et~al.}(2011)]{Cao2011}
Cao, S., Liang, N., \& Zhu, Z.-H. 2011, MNRAS, 416, 1099

\bibitem[Carloni {et~al.}(2024)]{Carloni2024}
Carloni, Y., Luongo, O., \& Muccino, M. 2024, arXiv:2404.12068

\bibitem[Carr {et~al.}(2022)]{Carr2022}
Carr, A., Davis, T.~M., Scolnic, D., \& et~al. 2022, PASA, 39, e046

\bibitem[Carroll {et~al.}(1992)]{Carroll1992}
Carroll, S.~M., Press, W.~H., \& Turner, E.~L. 1992, ARA\&A, 30, 499

\bibitem[Chen {et~al.}(2010)]{Chen2010}
Chen, Y., Zhu, Z.-H., Alcaniz, J.~S., \& Gong, Y. 2010, ApJ, 711, 439

\bibitem[Cheng {et~al.}(2021)]{Cheng2021}
Cheng, W., He, Y., Diao, J.-W., {et~al.} 2021, JHEP, 124

\bibitem[Chevallier \& Polarski(2001)]{Chevallier2001}
Chevallier, M., \& Polarski, D. 2001, IJMPD, 10, 213

\bibitem[Chiba(2002)]{Chiba2002}
Chiba, T. 2002, Phys. Rev. D, 66, 063514

\bibitem[Chimento {et~al.}(2013)]{Chimento2013}
Chimento, L.~P., Forte, M., \& Richarte, M.~G. 2013, Mod. Phys. Lett. A, 28,
  1250235

\bibitem[Chimento {et~al.}(2003)]{Chimento2003}
Chimento, L.~P., Jakubi, A.~S., Pavon, D., \& Zimdahl, W. 2003, Phys. Rev. D,
  67, 083513

\bibitem[Colg\'ain {et~al.}(2024{\natexlab{a}})]{Colgain2024a}
Colg\'ain, E.~O., Dainotti, M.~G., Capozziello, S., \& et~al.
  2024{\natexlab{a}}, arXiv:2404.08633

\bibitem[Colg\'ain {et~al.}(2024{\natexlab{b}})]{Colgain2024c}
Colg\'ain, E.~O., Pourojaghi, S., \& Sheikh-Jabbari, M.~M. 2024{\natexlab{b}},
  arXiv:2406.06389

\bibitem[Colg\'ain {et~al.}(2024{\natexlab{c}})]{Colgain2024b}
Colg\'ain, E.~O., Sheikh-Jabbari, M.~M., \& Yin, L. 2024{\natexlab{c}},
  arXiv:2405.19953

\bibitem[Collaboration {et~al.}(2024)]{DESI2024}
Collaboration, D., Adame, A.~G., Aguilar, J., Ahlen, S., \& et~al. 2024

\bibitem[Collaboration {et~al.}(2016)]{Planck2016}
Collaboration, P., Ade, P. A.~R., Aghanim, N., \& et~al. 2016, A\&A, 594, A13

\bibitem[Collaboration {et~al.}(2020)]{Planck2020}
Collaboration, P., Aghanim, N., Akrami, Y., \& et~al. 2020, A\&A, 641, A6

\bibitem[Cucchiara {et~al.}(2011)]{Cucchiara2011}
Cucchiara, A., Levan, A.~J., Fox, D.~B., \& et~al. 2011, ApJ, 736, 7

\bibitem[Cui \& Zhang(2010)]{CuiZhang2010}
Cui, J., \& Zhang, X. 2010, Phys. Lett. B, 690, 233

\bibitem[Dahiya \& Jain(2023)]{Dahiya2023}
Dahiya, D., \& Jain, D. 2023, Res. Astron. Astrophys., 23, 095001

\bibitem[Dai {et~al.}(2004)]{Dai2004}
Dai, Z.~G., Liang, E.~W., \& Xu, D. 2004, ApJ, 612, L101

\bibitem[Dainotti \& Del~Vecchio(2017)]{Dainotti2017}
Dainotti, M., \& Del~Vecchio, R. 2017, New Astronomy Reviews, 77, 23

\bibitem[Dainotti \& Amati(2018)]{Dainotti2018}
Dainotti, M.~G., \& Amati, L. 2018, PASP, 130, 051001

\bibitem[Dainotti {et~al.}(2023)]{Dainotti2023}
Dainotti, M.~G., Levine, D., Fraija, N., \& et~al. 2023, Galaxies, 11, 25

\bibitem[Dalal {et~al.}(2001)]{Dalal2001}
Dalal, N., Abazajian, K., Jenkins, E., \& Manohar, A.~V. 2001, Phys. Rev.
  Lett., 87, 141302

\bibitem[Demianski {et~al.}(2017{\natexlab{a}})]{Demianski2017a}
Demianski, M., Piedipalumbo, E., Sawant, D., \& et~al. 2017{\natexlab{a}},
  A\&A, 598, A112

\bibitem[Demianski {et~al.}(2017{\natexlab{b}})]{Demianski2017b}
Demianski, M., Piedipalumbo, E., Sawant, D., \& et~al. 2017{\natexlab{b}},
  A\&A, 598, A113

\bibitem[Di~Valentino {et~al.}(2017)]{Valentino2017}
Di~Valentino, E., Melchiorri, A., \& Mena, O. 2017, Phys. Rev. D, 96, 043503

\bibitem[Dirirsa {et~al.}(2019)]{Dirirsa2019}
Dirirsa, F.~F., Razzaque, S., Piron, F., \& et~al. 2019, ApJ, 887, 13

\bibitem[Eisenstein {et~al.}(2005)]{Eisenstein2005}
Eisenstein, D.~J., Zehavi, I., Hogg, D.~W., \& et~al. 2005, ApJ, 633, 560

\bibitem[Feng {et~al.}(2005)]{Feng2005}
Feng, B., Wang, X.~L., \& Zhang, X.~M. 2005, Phys. Lett. B, 607, 35

\bibitem[Foreman-Mackey {et~al.}(2013)]{ForemanMackey2013}
Foreman-Mackey, D., Hogg, D.~W., Lang, D., \& Goodman, J. 2013, PASP, 125, 306

\bibitem[Friaça {et~al.}(2005)]{Frica2005}
Friaça, A. C.~S., Alcaniz, J.~S., \& Lima, J. A.~S. 2005, MNRAS, 362, 1295

\bibitem[Gonzalez {et~al.}(2019)]{Goswami2019}
Gonzalez, G.~K., Pradhan, A., \& Beesham, A. 2019, Pramana – J. Phys., 93, 89

\bibitem[Gonzalez {et~al.}(2018)]{Gonzalez2018}
Gonzalez, J.~E., Silva, H. H.~B., Silva, R., \& Alcaniz, J.~S. 2018, EPJC, 78,
  730

\bibitem[Guo {et~al.}(2005)]{Guo2005}
Guo, Z.~K., Cai, R.~G., \& Zhang, Y.~Z. 2005, JCAP, 05, 002

\bibitem[Guo {et~al.}(2007)]{Guo2007}
Guo, Z.-K., Ohta, N., \& Tsujikawa, S. 2007, Phys. Rev. D, 76, 023508

\bibitem[Guo \& Zhang(2005)]{Guo2005b}
Guo, Z.~K., \& Zhang, Y.~Z. 2005, Phys. Rev. D, 71, 023501

\bibitem[Halder {et~al.}(2024)]{Halder2024}
Halder, S., de~Haro, J., Saha, T., \& Pan, S. 2024, arXiv: 2403.01397

\bibitem[Hasinger {et~al.}(2002)]{Hasinger2002}
Hasinger, G., Schartel, N., \& Komossa, S. 2002, ApJ, 573, L77

\bibitem[Hicken {et~al.}(2009)]{Hicken2009}
Hicken, M., Wood-Vasey, W.~M., Blondin, S., \& et~al. 2009, ApJ, 700, 1097

\bibitem[Hou {et~al.}(2023)]{Hou2023}
Hou, W.~T., Qi, J.~Z., Han, T., \& et~al. 2023, JCAP, 05, 017

\bibitem[Hu \& Wang(2023)]{HuWang2023}
Hu, J.~P., \& Wang, F.~Y. 2023, Universe, 9, 94

\bibitem[Hu {et~al.}(2021)]{Hu2021}
Hu, J.~P., Wang, F.~Y., \& Dai, Z.~G. 2021, MNRAS, 507, 730

\bibitem[Hu {et~al.}(2024)]{Hu2024}
Hu, J.~P., Wang, Y.~Y., Hu, J., \& Wang, F.~Y. 2024, A\&A, 681, A88

\bibitem[Jain \& Dev(2006)]{Jain2006}
Jain, D., \& Dev, A. 2006, Phys. Lett. B, 633, 436

\bibitem[Khadka {et~al.}(2021)]{Khadka2021}
Khadka, N., Luongo, O., Muccino, M., \& Ratra, B. 2021, JCAP, 09, 042

\bibitem[Komatsu {et~al.}(2009)]{Komatsu2009}
Komatsu, E., Dunkley, J., Nolta, M.~R., \& et~al. 2009, ApJS, 180, 330

\bibitem[Komatsu {et~al.}(2011)]{Komatsu2011}
Komatsu, E., Smith, K.~M., Dunkley, J., \& et~al. 2011, ApJS, 192, 18

\bibitem[Krisciunas {et~al.}(2017)]{Krisciunas2017}
Krisciunas, K., Contreras, C., Burns, C.~R., \& et~al. 2017, AJ, 154, 211

\bibitem[Kumar \& Nunes(2016)]{Kumar2016}
Kumar, S., \& Nunes, R.~C. 2016, Phys. Rev. D, 94, 123511

\bibitem[Kumar \& Nunes(2017)]{Kumar2017}
Kumar, S., \& Nunes, R.~C. 2017, Phys. Rev. D, 96, 103511

\bibitem[Kumar {et~al.}(2019)]{Kumar2019}
Kumar, S., Nunes, R.~C., \& Yadav, S.~K. 2019, Eur. Phys. J. C, 79, 7

\bibitem[Li {et~al.}(2024)]{Li2024}
Li, T.-N., Jin, S.-J., Li, H.-L., Zhang, J.-F., \& Zhang, X. 2024, ApJ, 963, 52

\bibitem[Li {et~al.}(2014)]{Li2014}
Li, Y.~H., Zhang, J.~F., \& Zhang, X. 2014, Phys. Rev. D, 90, 063005

\bibitem[Li \& Zhang(2023)]{LiZhang2023}
Li, Y.~H., \& Zhang, X. 2023, JCAP, 09, 046

\bibitem[Li {et~al.}(2023)]{Li2023}
Li, Z.~H., Zhang, B., \& Liang, N. 2023, MNRAS, 521, 4406

\bibitem[Liang {et~al.}(2022)]{Liang2022}
Liang, N., Li, Z., Xie, X., \& Wu, P. 2022, ApJ, 941, 84

\bibitem[Liang {et~al.}(2010)]{Liang2010}
Liang, N., Wu, P.~X., \& Zhang, S.~N. 2010, Phys. Rev. D, 81, 083518

\bibitem[Liang {et~al.}(2008)]{Liang2008}
Liang, N., Xiao, W.~K., Liu, Y., \& Zhang, S.~N. 2008, ApJ, 685, 354

\bibitem[Liang {et~al.}(2011)]{Liang2011}
Liang, N., Xu, L., \& Zhu, Z.~H. 2011, A\&A, 527, A11

\bibitem[Lima \& Alcaniz(2000)]{Lima2000}
Lima, J. A.~S., \& Alcaniz, J.~S. 2000, MNRAS, 317, 893

\bibitem[Linder(2003)]{Linder2003}
Linder, E.~V. 2003, Phys. Rev. Lett., 90, 091301

\bibitem[Liu {et~al.}(2022)]{Liu2022}
Liu, Y., Liang, N., Xie, X., \& et~al. 2022, ApJ, 935, 7

\bibitem[Luongo \& Muccino(2020)]{Luongo2020}
Luongo, O., \& Muccino, M. 2020, A\&A, 641, A174

\bibitem[Luongo \& Muccino(2021{\natexlab{a}})]{Luongo2021a}
Luongo, O., \& Muccino, M. 2021{\natexlab{a}}, Galaxy, 9, 77

\bibitem[Luongo \& Muccino(2021{\natexlab{b}})]{Luongo2021b}
Luongo, O., \& Muccino, M. 2021{\natexlab{b}}, MNRAS, 503, 4581

\bibitem[Luongo \& Muccino(2024)]{Luongo2024}
Luongo, O., \& Muccino, M. 2024, arXiv:2024.07070

\bibitem[Montiel {et~al.}(2021)]{Montiel2021}
Montiel, A., Cabrera, J.~I., \& Hidalgo, J.~C. 2021, MNRAS, 501, 3515

\bibitem[Moresco {et~al.}(2022)]{Moresco2022}
Moresco, M., Amati, L., Amendola, L., \& et~al. 2022, Living Reviews in
  Relativity, 25, 6

\bibitem[Pan {et~al.}(2019)]{Pan2019}
Pan, S., Yang, W., Di~Valentino, E., Saridakis, E.~N., \& Chakraborty, S. 2019,
  Phys. Rev. D, 100, 103520

\bibitem[Pan {et~al.}(2020)]{Pan2020}
Pan, S., Yang, W., \& Paliathanasis, A. 2020, MNRAS, 493, 3114

\bibitem[Pavon {et~al.}(2004)]{Pavon2004}
Pavon, D., Sen, S., \& Zimdahl, W. 2004, JCAP, 05, 009

\bibitem[Percival {et~al.}(2010)]{Percival2010}
Percival, W.~J., Reid, B.~A., Eisenstein, D.~J., \& et~al. 2010, MNRAS, 401,
  2148

\bibitem[Perivolaropoulos \& Skara(2022)]{Perivolaropoulos2022}
Perivolaropoulos, L., \& Skara, F. 2022, New Astron. Rev., 95, 101659

\bibitem[Perlmutter {et~al.}(1999)]{Perlmutter1999}
Perlmutter, S., Aldering, G., Goldhaber, G., \& et~al. 1999, ApJ, 517, 565

\bibitem[Ratra \& Peebles(1988)]{Ratra1988}
Ratra, B., \& Peebles, P. J.~E. 1988, Phys. Rev. D, 37, 3406

\bibitem[Riess {et~al.}(2022{\natexlab{a}})]{Riess2022b}
Riess, A.~G., Breuval, L., Yuan, W., \& et~al. 2022{\natexlab{a}}, ApJ, 938, 36

\bibitem[Riess {et~al.}(2019)]{Riess2019}
Riess, A.~G., Casertano, S., Yuan, W., \& et~al. 2019, ApJ, 876, 85

\bibitem[Riess {et~al.}(1998)]{Riess1998}
Riess, A.~G., Filippenko, A.~V., Challis, P., \& et~al. 1998, AJ, 116, 1009

\bibitem[Riess {et~al.}(1999)]{Riess1999}
Riess, A.~G., Kirshner, R.~P., Schmidt, B.~P., \& et~al. 1999, AJ, 117, 707

\bibitem[Riess {et~al.}(2016)]{Riess2006}
Riess, A.~G., Macri, L.~M., Hoffmann, S.~L., \& et~al. 2016, ApJ, 826, 56

\bibitem[Riess {et~al.}(2018)]{Riess2018}
Riess, A.~G., Rodney, S.~A., Scolnic, D.~M., \& et~al. 2018, ApJ, 853, 126

\bibitem[Riess {et~al.}(2022{\natexlab{b}})]{Riess2022a}
Riess, A.~G., Yuan, W., Macri, L.~M., \& et~al. 2022{\natexlab{b}}, ApJL, 934,
  L7

\bibitem[Sako {et~al.}(2018)]{Sako2018}
Sako, M., Bassett, B., Becker, A.~C., \& et~al. 2018, PASP, 130, 064002

\bibitem[Schaefer(2007)]{Schaefer2007}
Schaefer, B.~E. 2007, ApJ, 660, 16

\bibitem[Schwarz(1978)]{Schwarz1978}
Schwarz, G. 1978, Ann. Statist., 6, 461

\bibitem[Scolnic {et~al.}(2022)]{Scolnic2022}
Scolnic, D., Brout, D., Carr, A., \& et~al. 2022, ApJ, 938, 113

\bibitem[Scolnic {et~al.}(2018)]{Scolnic2018}
Scolnic, D.~M., Jones, D.~O., Rest, A., \& et~al. 2018, ApJ, 859, 101

\bibitem[Sethi {et~al.}(2005)]{Sethi2005}
Sethi, G., Dev, A., \& Jain, D. 2005, Phys. Lett. B, 624, 135

\bibitem[Spergel {et~al.}(2007)]{Spergel2007}
Spergel, D.~N., Bean, R., Dore, O., \& et~al. 2007, ApJS, 170, 377

\bibitem[Spergel {et~al.}(2003)]{Spergel2003}
Spergel, D.~N., Verde, L., Peiris, H.~V., \& et~al. 2003, ApJS, 148, 175

\bibitem[Tegmark {et~al.}(2004)]{Tegmark2004}
Tegmark, M., Blanton, M.~R., Strauss, M.~A., \& et~al. 2004, ApJ, 606, 702

\bibitem[Tripp(1998)]{Tripp1998}
Tripp, R. 1998, A\&A, 331, 815

\bibitem[von Marttens {et~al.}(2019)]{vonMarttens2019}
von Marttens, R., Casarini, L., Mota, D.~F., \& Zimdahl, W. 2019, Physics of
  the Dark Universe, 23, 100248

\bibitem[Wang(2022)]{Wang2022}
Wang, D. 2022, Phys. Rev. D, 106, 063515

\bibitem[Wang {et~al.}(2015)]{Wang2015}
Wang, F.~Y., Dai, Z.~G., \& Liang, E.~W. 2015, Nar, 67, 1

\bibitem[Wang {et~al.}(2022)]{WH2022}
Wang, F.~Y., Hu, J.~P., Zhang, G.~Q., \& Dai, Z.~G. 2022, ApJ, 924, 97

\bibitem[Wang {et~al.}(2024)]{WLL2024}
Wang, G., Li, X., \& Liang, N. 2024, APSS, 369, 74

\bibitem[Wang \& Liang(2024)]{WL2024}
Wang, H., \& Liang, N. 2024, MNRAS, 533, 743

\bibitem[Wang {et~al.}(2016)]{Wang2016}
Wang, J.~S., Wang, F.~Y., Cheng, K.~S., \& Dai, Z.~G. 2016, A\&A, 585, A68

\bibitem[Wang {et~al.}(2010)]{Wang2010}
Wang, S., Li, X.-D., \& Li, M. 2010, Phys. Rev. D, 82, 103006

\bibitem[Wang \& Zhang(2008)]{WangZhang2008}
Wang, S., \& Zhang, Y. 2008, Phys. Lett. B, 669, 201

\bibitem[Wei(2010{\natexlab{a}})]{Wei2010a}
Wei, H. 2010{\natexlab{a}}, JCAP, 08, 020

\bibitem[Wei(2010{\natexlab{b}})]{Wei2010b}
Wei, H. 2010{\natexlab{b}}, Phys. Lett. B, 691, 173

\bibitem[Wei \& Zhang(2007{\natexlab{a}})]{WeiZhang2007a}
Wei, H., \& Zhang, S.~N. 2007{\natexlab{a}}, Phys. Lett. B, 654, 139

\bibitem[Wei \& Zhang(2007{\natexlab{b}})]{WeiZhang2007b}
Wei, H., \& Zhang, S.~N. 2007{\natexlab{b}}, Phys. Lett. B, 644, 7

\bibitem[Wei \& Zhang(2007{\natexlab{c}})]{WeiZhang2007c}
Wei, H., \& Zhang, S.~N. 2007{\natexlab{c}}, Phys. Rev. D, 76, 063003

\bibitem[Weinberg(1989)]{Weinberg1989}
Weinberg, S. 1989, Rev. Mod. Phys., 61, 1

\bibitem[Yan {et~al.}(2015)]{Yan2015}
Yan, X.-P., Liu, D.-Z., \& Wei, H. 2015, Phys. Lett. B, 742, 149

\bibitem[Yang \& Zhang(2010)]{YangZhang2010}
Yang, R.-J., \& Zhang, S.~N. 2010, MNRAS, 407, 1835

\bibitem[Yang {et~al.}(2018)]{Yang2018}
Yang, W., Pan, S., Di~Valentino, E., {et~al.} 2018, JCAP, 09, 019

\bibitem[Yang {et~al.}(2020)]{Yang2020}
Yang, W., Pan, S., Nunes, R.~C., \& Mota, D.~F. 2020, JCAP, 04, 008

\bibitem[Yu \& Wang(2014)]{YuWang2014}
Yu, H., \& Wang, F.~Y. 2014, Eur. Phys. J. C, 74, 3090

\bibitem[Zarandi \& Ebrahimi(2022)]{ZarandiEbrahimi2022}
Zarandi, H. R.~M., \& Ebrahimi, E. 2022, Int. J. Mod. Phys. D, 31, 2250121

\bibitem[Zhao {et~al.}(2023)]{Zhao2023}
Zhao, Z.-W., Wang, L.-F., Zhang, J.-G., \& et~al. 2023, JCAP, 04, 022

\bibitem[Zheng {et~al.}(2022)]{Zheng2022}
Zheng, J., Chen, Y., Xu, T., \& Zhu, Z.~H. 2022, Eur. Phys. J. Plus, 137, 4

\bibitem[Zimdahl {et~al.}(2001)]{Zimdahl2001}
Zimdahl, W., Pavon, D., \& Chimento, L.~P. 2001, Phys. Lett. B, 521, 133

\end{thebibliography}


\end{document}